\documentclass{article}

    \PassOptionsToPackage{numbers, compress}{natbib}

 \usepackage[main, final]{neurips_2026}

\usepackage{hyperref}
\usepackage{url}
\usepackage{enumitem}

\usepackage[utf8]{inputenc} 
\usepackage[T1]{fontenc}    
\usepackage{hyperref}       
\usepackage{url}            
\usepackage{booktabs}       
\usepackage{amsfonts}       
\usepackage{nicefrac}       
\usepackage{microtype}      
\usepackage{xcolor}         
\usepackage{graphicx}%
\usepackage{float}
\usepackage{amsmath}
\usepackage{caption, graphicx}
\usepackage{wrapfig}

\newcommand{\SHIFT}{\emph{socio-situational shift}}

\title{Unifying Partner and Environment Diversity to Improve Human-AI Coordination}

\author{%
  Patrick Q. Zhang \\
  \texttt{pqz317@uw.edu} \\
  \And
  Yancheng Liang \\
  \texttt{yancheng@cs.washington.edu} \\
  \AND
  Adrienne Fairhall \\
  \texttt{fairhall@uw.edu} \\
  \And
  Natasha Jaques \\
  \texttt{nj@cs.washington.edu} \\
}

\begin{document}

\maketitle

\begin{abstract}

Real-world applications of modern AI systems require generalization to both unseen environments and novel human users. The field of zero-shot coordination (ZSC) has focused primarily on deploying AI to cooperate with novel partners in otherwise familiar environments. Popular population-based methods achieve this by partnering the agent with a diversity of simulated human users during training in a fixed environment, but does not consider the more realistic scenario where both the users and the environment are unseen during training. Recently, training across environments has been explored in training a single self-play agent, which shows better generalization on both novel human users and unseen environments. In this work, we ask whether jointly modeling partner diversity and environment diversity yields stronger coordination with novel partners in novel environments. We introduce Cross Environment Co-Play (CECP), a two-stage framework that first trains a population of self-play agents across a distribution of procedurally generated environments, and then trains a cooperator policy against this partner population. To further couple social and environmental reasoning, we add a next-state prediction module that encourages representations to encode both partner behavior and task structure. We evaluate CECP on both a neuroscience-inspired cooperative foraging task, and the popular ZSC benchmark Overcooked, under held-out partner and held-out environment conditions. CECP achieves the strongest performance among the compared baselines and attains state-of-the-art results under our evaluation protocol for human-AI collaboration. We analyze the model’s latent representation encoded by the next-state prediction module, showing it can not only improve the performance of the cooperative AI, but also provide more interpretable latents, demonstrated by the inference of partner skill levels. 
\footnote{See our website: 
\href{https://pqz317.github.io/cecp-project-site/}{https://pqz317.github.io/cecp-project-site/} and training
code: \href{https://github.com/pqz317/cecp}{https://github.com/pqz317/cecp}.}

\end{abstract}

\section{Introduction}

AI continues to make incredibly rapid progress, especially in digital domains. However, current AI progress looks very different from the idealized future many of us imagined. We may have solved the Navier-stokes problem, but a true household robot that helps with chores and alleviates manual labor remains the stuff of science fiction; we are still left doing our own dishes! To realize this vision of physically assistive AI, we need an AI agent that can assist many different people, in many different homes, across a wide range of tasks; a capability humans exercise effortlessly, but one that AI agents still struggle with. We define the joint adaptation to both environments and novel humans as \SHIFT. 


Zero-shot coordination (ZSC) ~\citep{carroll_utility_2020} provides a common framework for studying AI generalization across diverse human users. In typical ZSC benchmarks such as the cooperative cooking game Overcooked, agents are paired with unseen human partners ~\citep{strouse_collaborating_2022, zhao_maximum_2022, liang_learning_2024, chaudhary_improving_2025, yan_efficient_2023} on the same environment that is used for training. Currently, over 700 papers have been written on ZSC using Overcooked, and almost all of them focus purely on covering the space of possible partners, while neglecting the dimension of being able to cooperate on-the-fly with a new partner \emph{on a new environment with potentially new tasks or objectives}. Only a handful of papers ~\citep{jha_cross-environment_2025,ruhdorfer_overcooked_2025}, have focused on the idea of training agents across multiple different environments in a ZSC context. Intriguingly, \citet{jha_cross-environment_2025} found that merely training on diverse environments can lead to effective generalization to human partners, even when using self-play. However, as yet, no one has successfully explored combining both partner and environment diversity. We hypothesize that algorithms that can handle this high degree of both task and strategic complexity will be required for some of the most beneficial uses of AI, such as embodied cooperative assistants.



\begin{figure}[t]
  \centering
  \includegraphics[width=1\textwidth]{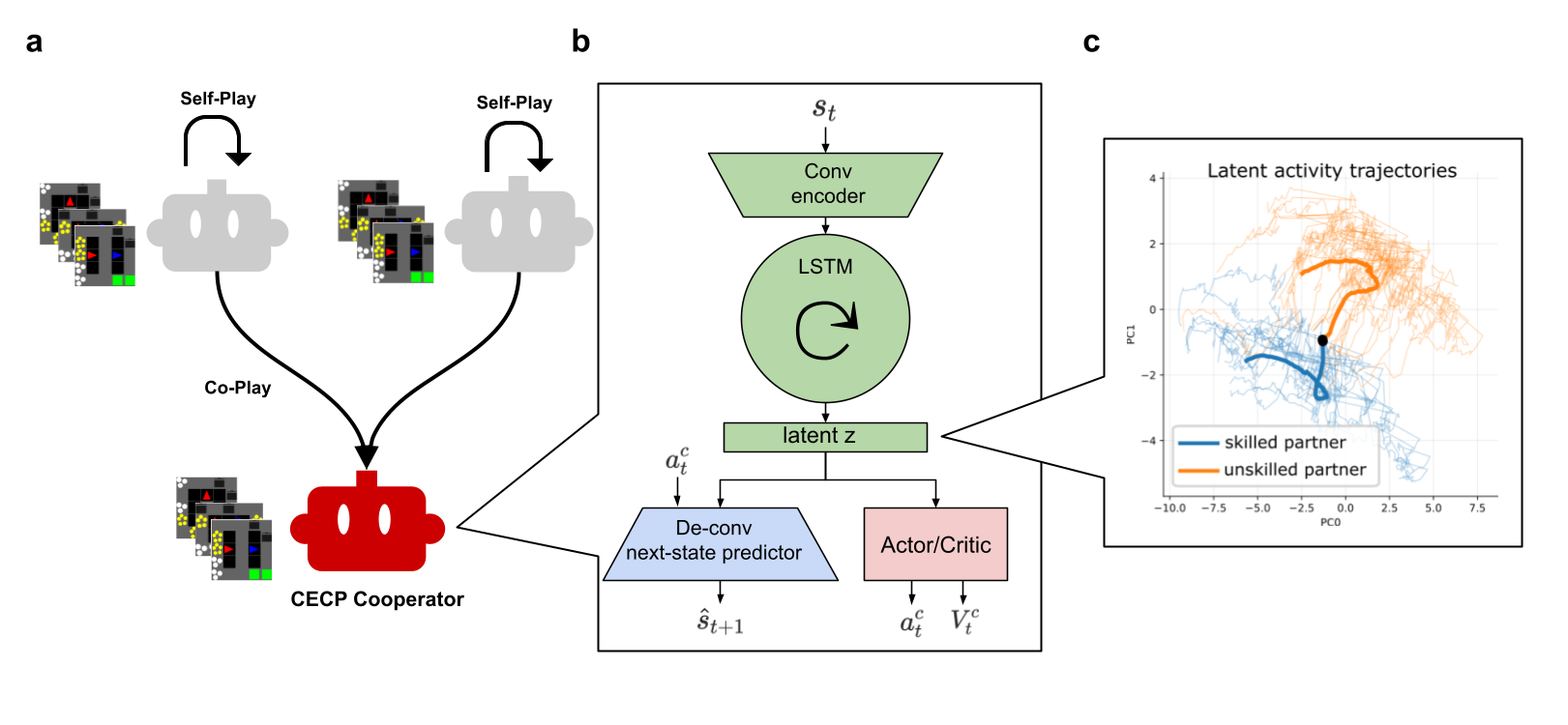}
    \caption{An overview of CECP. \textbf{a}, Training paradigm. Partner agents are first trained via a self-play on a distribution of environments. Then, the CECP cooperator trains by partnering with the frozen partner population, training also on a distribution of environments. \textbf{b,} Architecture of the cooperator agent. CECP is equipped with a next-state prediction objective (blue). \textbf{c,} Latent representation in the cooperator agent. CECP latents form representations of partner behavior, such as inferred partner skill level. }\label{intro_fig}    
\end{figure}


In this work, we study cooperative generalization under simultaneous \SHIFT. We introduce Cross-Environment Co-Play (CECP), a meta-reinforcement learning approach designed to unify these two axes. The training of CECP proceeds in two stages (Figure \ref{intro_fig}a). \textit{Training environment and partner setup}: A population of agents is trained via self-play over a procedurally generated distribution of environments, producing a diverse set of policies (the partner pool) grounded in varied dynamics. \textit{Training-stage for meta-RL}: A cooperator policy is trained against a frozen population sampled from the partner pool, exposing it to heterogeneous partner behaviors while retaining broad environmental variation. \textit{Evaluation}: The cooperator will be evaluated against novel partners (for example, human players) in unseen environments.

Training along both partner behavior diversity and environment variation simultaneously makes robust representation learning especially challenging due to high non-stationarity. To address this, we incorporate a next-state prediction auxiliary objective for the cooperator agent (Figure \ref{intro_fig}b, blue). The intuition is that in cooperative settings, predicting state transitions requires modeling both environment dynamics and partner actions, encouraging the agent to build representations that capture both. Enabling the cooperator agent to implicitly predict the partners' actions may also lead to faster inference and better adaptation to the partner's coordination behaviors. With the ablation study, we show that the auxiliary training objective consistently improves generalization performance across the joint diversity training regime.


We evaluate CECP in a cooperative foraging task and Overcooked domains under protocols that include both held-out partners and held-out environments. Across these settings, CECP consistently outperforms strong baselines, demonstrating improved generalization under combined social and environmental shifts. Furthermore, we conduct human experiments and show that under real-time evaluation against novel human players, CECP reached the highest coordination performance with human participants, and participants rated CECP agents as most likely to adapt to their actions. In our interpretability analysis, we find that including the next state predictive objective and partner diversity encourages the model to form representations of partner actions and skill level (Figure \ref{intro_fig}c). 

Our contributions are:
\begin{itemize}[leftmargin=*]
    \item We propose CECP, a two-stage training paradigm that combines partner diversity and environment diversity in a unified setting.
    \item We empirically demonstrate improved performance on both the cooperative foraging tasks and Overcooked, a popular ZSC benchmark, under simultaneous \SHIFT.
    \item We find that CECP agents form representations that predict partner actions and infer partner skill level, providing a useful tool for insightful and interpretable analysis.
\end{itemize}

\section{Related Work}

\textbf{Zero-shot Coordination (ZSC).} \textbf{Self-play (SP)} describes the training method in which an agent learns from repeated games with itself. Despite the great success in competitive tasks like Starcraft and Go ~\citep{vinyals_grandmaster_2019, silver_mastering_2017}, in collaborative settings, however, SP agents are known to struggle in cooperative tasks, in that they often fail to coordinate with novel partners ~\citep{strouse_collaborating_2022}. SP agents tend to learn specific coordination patterns that are work only when playing with themselves. Extensive works ~\citep{zhao_maximum_2022,liang_learning_2024, lupu_trajectory_2021, sarkar_diverse_2023, chaudhary_improving_2025} use \textbf{Population-based training (PBT)} methods ~\citep{jaderberg_population_2017,carroll_utility_2020} to build a diverse population of training partners, aiming to cover the test-time behaviors of novel human players. While the above methods are typically trained and evaluated on the same single environments, our work further explores injecting environment diversity in both the stage of creating the partner population and the stage of training the cooperator against the population. 

\textbf{Meta-reinforcement learning (Meta-RL)} aims to learn policies that adapt rapidly to novel tasks by training on a distribution of tasks or environments, rather than a single fixed setting ~\citep{beck_survey_2023,yu_meta-world_2021,duan_rl2_2016,wang_learning_2017,zintgraf_varibad_2021}. Similar methods have also been extended to multi-agent settings ~\citep{zintgraf_deep_2022, raileanu_modeling_2018, rabinowitz_machine_2018, al-shedivat_continuous_2018, zhang_learning_2023}.  Recently in ZSC, \citet{jha_cross-environment_2025}  demonstrated that leveraging meta-RL to train a single self-play agent across a distribution of procedurally generated environments improves the agent's ability to also coordinate with unseen partners. However, when \citet{jha_cross-environment_2025} attempted to combine partner and environment diversity, they were not able to achieve good performance with standard models, motivating our investigation into what techniques would be needed to handle this highly complex training distribution

\textbf{Predictive Auxiliary Objective}
Recent work suggests that predictive modules can improve partner adaptation in multi-agent coordination. In particular, explicit partner-action modeling has proven useful for inferring partner intent ~\citep{rabinowitz_machine_2018} and conditioning ego behavior on predicted teammate actions ~\citep{zintgraf_deep_2022,yan_efficient_2023,ni_theory_2026}.  Beyond direct partner prediction, auxiliary predictive objectives have also improved representation learning in cooperative multi-agent settings ~\citep{feng_joint-predictive_2022,huang_predictive_2025}. In meta-RL and under partial observability, predictive objectives have similarly been shown to improve Bayes-adaptive belief representations ~\citep{zintgraf_varibad_2021,kuo_predictive_2025}. Our work is inspired by \citet{ndousse_emergent_2021}, which demonstrated that a next-state predictive auxiliary objective enabled social learning in a multi-agent context, with the reasoning that accurate next-state prediction must implicitly capture both partner action and environment transitions.

\section{Preliminaries}

A two-player cooperative Markov game is a tuple $\langle \mathcal{S}, \mathcal{A}, \mathcal{T}, R, \rho_0, H \rangle$, where $\mathcal{S}$ is the state space, $\mathcal{A}$ is the shared action space, $\mathcal{T}: \mathcal{S} \times \mathcal{A} \times \mathcal{A} \to \Delta(\mathcal{S})$ is the transition function, $R: \mathcal{S} \times \mathcal{A} \times \mathcal{A} \to \mathbb{R}$ is the shared reward function, $\rho_0 \in \Delta(\mathcal{S})$ is the initial state distribution, and $H$ is the horizon. For a cooperator policy $\pi_c$, and a partner policy $\pi_p$, we define the expected return as
\begin{equation}
S(\pi_c, \pi_p) = \mathbb{E}_{\substack{ s_0 \sim \rho_0, s_{t+1} \sim \mathcal{T}, \\ a_t^c \sim \pi_c, \, a_t^p \sim \pi_p}} \left[ \sum_{t=0}^{H} R(s_t, a_t^c, a_t^p) \right].
\end{equation}

\textbf{Population-based training (PBT): }
We assume a distribution over partner policies $\pi_p \sim \mathcal{P}$, representing the variation in strategy, convention, and skill that the cooperator may encounter at deployment. During training, the cooperator's objective is: 
\begin{equation}
J_{\text{PBT}}(\pi_c) = \mathbb{E}_{\pi_p \sim \mathcal{P}} \left[ S(\pi_c, \pi_p) \right].
\end{equation}

\textbf{Meta-RL:}
We consider a distribution $\mathcal{M}$ over Markov games following previous work ~\citep{jha_cross-environment_2025}. Each $m \sim \mathcal{M}$ shares the same $\mathcal{S}, \mathcal{A}, \mathcal{T}, R, H$, but has a distinct initial state distribution $\rho_0^m$. Writing $S(\pi_c, \pi_p; m)$ for the return under game $m$, the training objective for self-play (CEC, \cite{jha_cross-environment_2025}) is defined as: 

\begin{equation}
J_{\text{CEC}}(\pi_c) = \mathbb{E}_{m \sim \mathcal{M}} \left[ S(\pi_c, \pi_c; m) \right].
\end{equation}



\section{CECP (Cross-Environment Co-Play)}

CECP trains a cooperator agent that generalizes over both tasks and partners. 
Training proceeds in two stages.

\textbf{Stage 1: Partner population training across environments.}
We train $N$ partner agents independently via self-play, under a distribution of environments generated via procedural generation ~\citep{jha_cross-environment_2025}. We use a JAX implementation of Overcooked ~\citep{rutherford_jaxmarl_2024} to enable 4096 parallel environment simulations and around 30,000 frames/s for training on a NVIDIA A100 GPU. Each self-play partner $\pi_p^{(i)}$ is obtained by:
\begin{equation}
    \pi_p^{(i)} = \arg\max_{\pi}\ \mathbb{E}_{m \sim \mathcal{M}}\left[S(\pi, \pi; m)\right],
    \quad i = 1, \ldots, N
    \label{eq:per_partner_cec}
\end{equation}

To further introduce diversity in partner skill level, we follow the FCP protocol ~\citep{strouse_collaborating_2022} agents, we vary the random seeds for each partner agent, and add a set of intermediate training checkpoints (early, middle, late) into the partner population $\mathcal{P}_{\mathcal{M}}$.

\textbf{Stage 2: Cooperator training against \SHIFT.}
A cooperator $\pi_c$ is then trained to maximize expected return jointly over 
the task and partner distributions:
\begin{equation}
    J_{\text{CECP}}(\pi_c) = \mathbb{E}_{m \sim \mathcal{M},\; \pi_p \sim 
    \mathcal{P}_{\mathcal{M}}}\!\left[S(\pi_c, \pi_p; m)\right]
    \label{eq:cecp_objective}
\end{equation}

\textbf{Auxiliary next-state prediction.}
To support representation learning under \SHIFT, 
we equip the cooperator with an auxiliary next-state prediction objective. The cooperator architecture consists of three components:
(1) An RNN encoder $q_{\phi}(z \mid \tau_{0:t})$ mapping observation histories to a latent representation $z$. (Figure \ref{intro_fig}b, green). (2) A prediction head $p_{\theta}(s_{t+1} \mid z, a_t^c)$ that decodes the next state from the latent and the cooperator's own action (Figure \ref{intro_fig}b, blue). (3) A policy head $\pi_{\psi}(a^c_t \mid z)$, conditioned on the latent $z$ and trained via PPO (Figure \ref{intro_fig}b, red).

The auxiliary prediction loss is:
\begin{equation}
\mathcal{L}_{\text{pred}}(\phi, \theta) 
    = \mathbb{E}_{(\tau_{0:t},\, a_t^c,\, s_{t+1})} 
      \left[ -\log p_{\theta}\!\left(s_{t+1} \mid q_{\phi}(\tau_{0:t}),\, a_t^c\right) \right].
\end{equation}
Because $s_{t+1}$ depends on the partner action $a_t^p$ and the 
task-specific dynamics $\mathcal{T}$, minimizing $\mathcal{L}_{\text{pred}}$ 
requires the encoder $q_\phi$ to capture information about both partner behavior 
and environment structure in $z$ ~\citep{ndousse_emergent_2021}. The full objective is:

\begin{equation}
    J_{\text{CECP-}\lambda}(\pi_c) = J_{\text{CECP}}(\pi_c) - \lambda\, \mathcal{L}_{\text{pred}}(\theta, \phi).
\end{equation}

\begin{wrapfigure}{H}{0.3\textwidth}
  \vspace{-1em}
  \centering
  \includegraphics[width=0.25\textwidth]{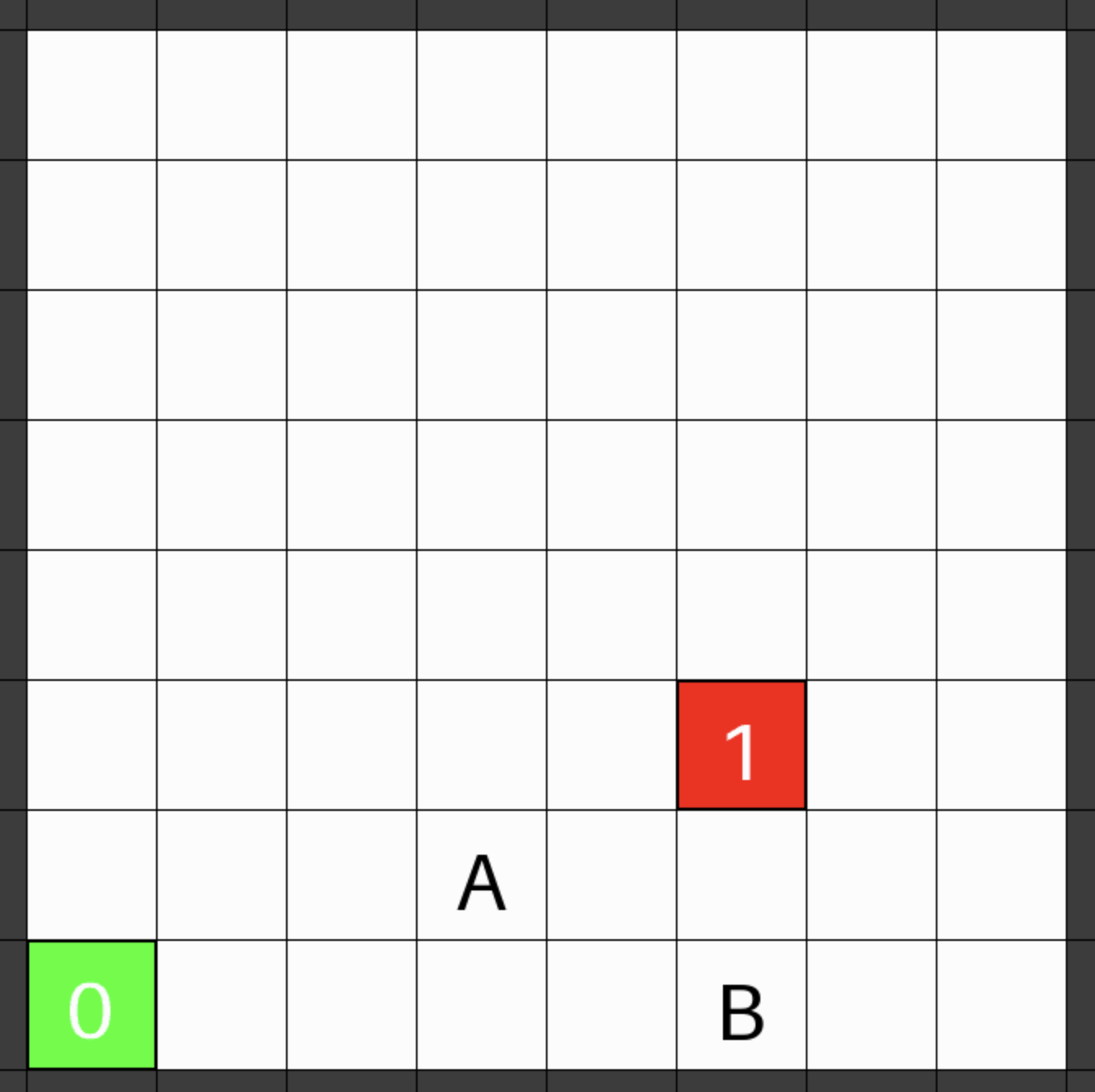}
  \caption{Cooperative Foraging environment. Both agents (0 and 1) must reach and stay on the same goal (A or B).} \label{coop_foraging_illustration}
  \vspace{-5em}
\end{wrapfigure}

\section{Experimental Setup}

We motivate our experiments with three questions related to our method: 

\textbf{Q1}: Can CECP leverage environment and partner diversity to generalize to novel environment/partner combinations?

\textbf{Q2}: Does our method obtain better performance than competitive baselines when coordinating against novel human users?

\textbf{Q3:} How do next state prediction and partner diversity shape representation learning and ZSC performance?

\subsection{Environments}

\textbf{Cooperative Foraging:} In this task, inspired by cooperation tasks used in neuroscience ~\citep{cheng_asymmetric_2025}, 2 agents are in a grid-world with 2 goal locations. To jointly receive reward, both agents must reach and stay on the same goal location. To generate a distribution of environments, we uniformly sample locations to place the two goals, and hold-out a set of goal locations for evaluation (Fig \ref{coop_foraging_illustration}). 

\begin{wrapfigure}{H}{0.35\textwidth}
  \centering
  \includegraphics[width=0.3\textwidth]{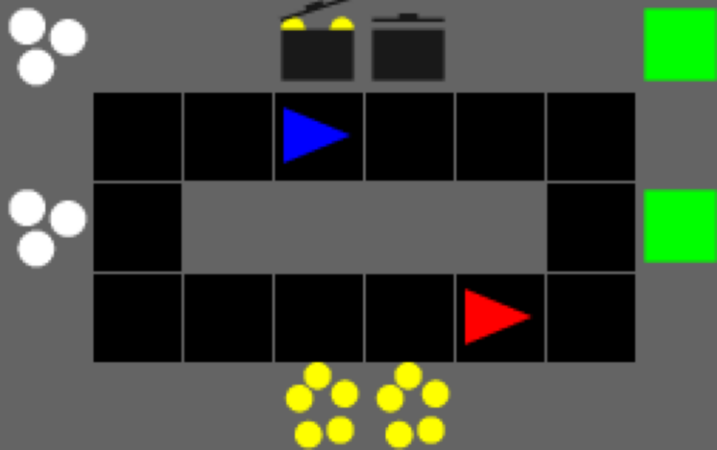}
  \caption{Visualization of the Counter Circuit layout, one of the original 5 Overcooked layouts. Two chefs (blue and red) need to cooperate together to make and deliver the soups efficiently.} \label{oc_counter_circuit}
\end{wrapfigure}

\textbf{Overcooked: } 
In the popular ZSC benchmark Overcooked ~\citep{carroll_utility_2020}, two players need to cooperate with each other to divide cooking responsibilities, and players jointly receive a reward by delivering completed dishes (Figure \ref{oc_counter_circuit}). 5 original layouts (Figure \ref{oc_original_5_layouts}, Appendix) are traditionally used to evaluate different coordination challenges. We evaluate our agents on the 5 original layouts, which are held out during training for CEC and CECP methods, and on an additional 100 procedurally generated layouts that are also held-out during training. 

\subsection{Baselines}
We evaluate our method, CECP, against existing baselines in the literature, particularly self-play (SP), fictitious co-play (FCP,  \cite{strouse_collaborating_2022}), and cross-environment cooperation (CEC, \cite{jha_cross-environment_2025}). For Overcooked evaluations, we also include: Maximum Entropy Population (MEP, \cite{zhao_maximum_2022}), a population-based method which enforces population diversity through a population entropy bonus; Cross-play optimized, Mixed-play enforced Diversity (CoMeDi, \cite{sarkar_diverse_2023}), another population-based method which enforces population diversity through mixed play; and Peer Adaptation method with Context-aware Exploration (PACE, \cite{ma_fast_2024}), a method which seeks to improve a cooperator by adding a peer-identification reward.

\subsection{Evaluation Criteria}

\textbf{Cross-play: } We evaluate coordination between independently trained agents drawn from the same training paradigm and architecture, but different initial seeds. Since agents with different initial seeds are never seen as partners during training in any of the methods, cross-play performance measures an agent's ability to coordinate with novel partners in a AI-AI coordination setting. 

\textbf{Human-AI coordination: } In Overcooked, we evaluate Human-AI ZSC by assessing how well an agent cooperates with both a behavior-cloned human proxy agent, and real human partners. We recruited 73 participants from the popular crowd work platform Prolific. Participants were compensated at an average rate of \$19.46/hr. All procedures were approved by our Institutional Review Board (IRB). Recruited human participants play against each agent for 200 timesteps each, and their performance is recorded along with questionnaire answers about coordination sentiment.

\section{Results}

\subsection{CECP generalizes to novel environment and partner settings in cooperative foraging}

\begin{figure}[H]
  \centering
  \includegraphics[width=0.6\textwidth]{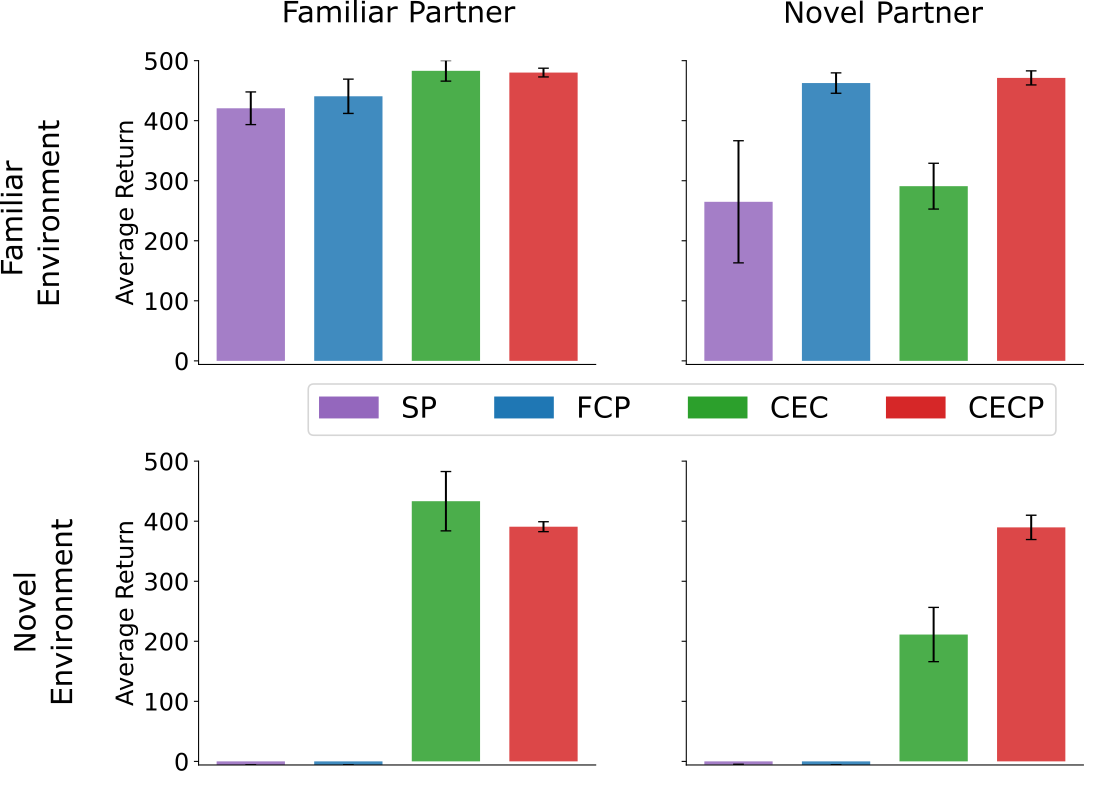}
  \caption{Performance of CECP along baseline methods in the Cooperative Foraging task across familiar/novel environment and partner settings. All previous methods, except CEC, which also adopts cross-environment training, fail on held-out environments. CECP shows additional robustness on \SHIFT.}
  \label{coop_foraging_grid}
\end{figure}

To investigate \textbf{Q1}, we evaluate CECP along with existing baselines in the Cooperative Foraging task. To test generalization to both novel environments and novel partners, we hold out a test set of goal locations, and a test set of partner agents for evaluation. Figure \ref{coop_foraging_grid} shows these results. While all methods perform well in the context of familiar partners and familiar environments, SP (purple) and FCP (blue) both fail to generalize when goal locations are different from their training environment. In comparison, CEC (green) generalizes to held-out goal locations well, but suffers a drop in performance when paired with a novel partner. Under the simultaneous \SHIFT\   (Figure 4, bottom right), CEC gets about half the performance of CECP (red). Our method is the only one that can simultaneously generalize to cooperate with new partners in new environments. 

The drop in CEC performance from familiar to novel partners suggests that in this setting, environment diversity alone does not prevent self-play agents from forming conventions, and motivates leveraging the diversity of conventions formed across a population of CEC agents to train a cooperator agent (CECP), to generalize to unseen partners.

\subsection{CECP performs better than baselines at zero-shot Human-AI coordination in Overcooked}

\begin{figure}[H]
  \centering
  \includegraphics[width=0.8\textwidth]{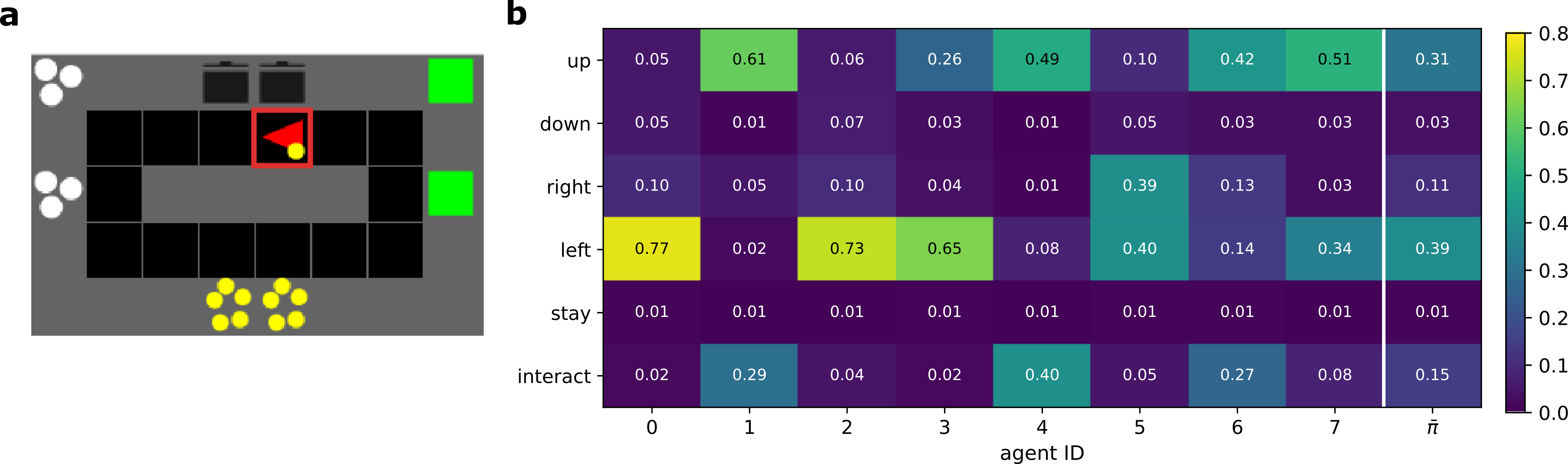}
  \caption{Example behavioral differences in a population of CEC agents, in a specific scenario in the Counter Circuit layout. \textbf{a, } an agent is holding an onion, facing left, under the rightward pot. \textbf{b, } averaged action distributions of CEC agents in the population, when faced with the scenario in \textbf{a}.}\label{cec_convention_counter_circuit}
\end{figure}

\begin{figure}[t]
  \centering
  \includegraphics[width=1\textwidth]{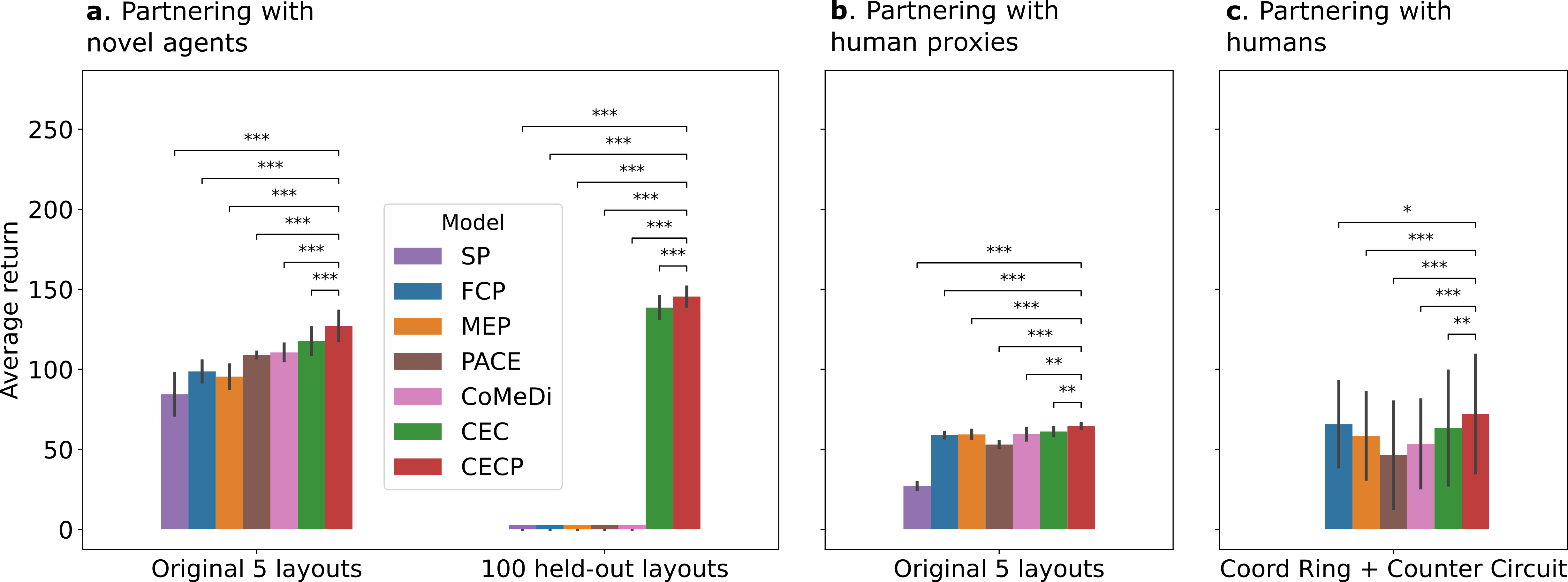}
  \caption{Performance on Overcooked layouts. \textbf{a,} Cross-play average returns of CECP and baselines across the 5 original Overcooked layouts. \textbf{b, } Average return of each method when paired with human proxy agents, trained via behavioral cloning (BC). \textbf{c,} Average return of each method when paired with humans. Significance assessed via two-sided Welch's t-test, ***: $p < 0.001$, **: $p < 0.01$, *: $p < 0.05$.}\label{avg_returns_combined}
\end{figure}

\begin{figure}[t]
  \centering
  \includegraphics[width=0.9\textwidth]{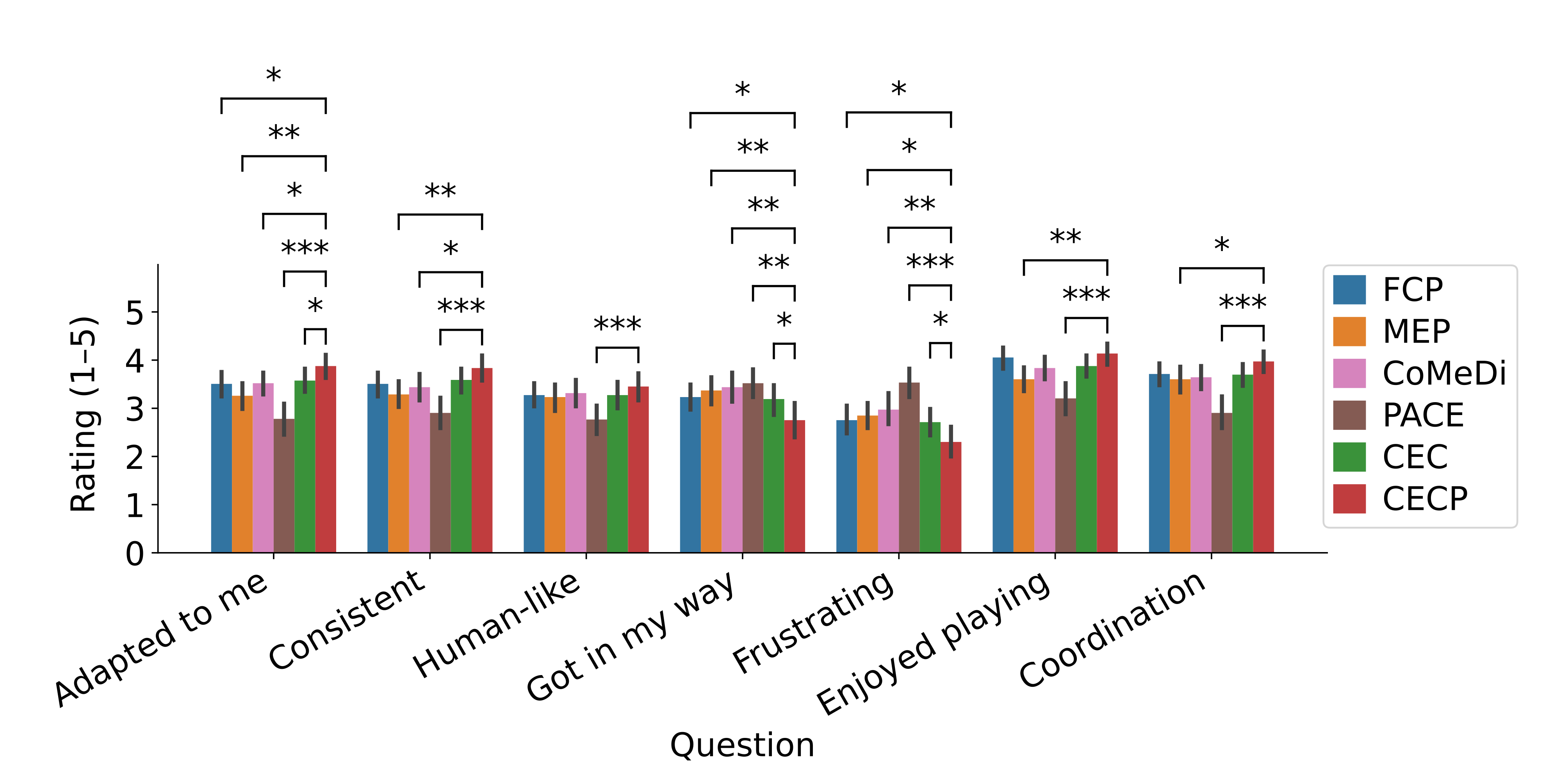}
  \caption{Average participant rating in response to survey questions (Likert scale, 1-5) for each method, significance tested with Paired Wilcoxon signed-rank tests, with *: $p < 0.05$, **: $p<0.01$, ***: $p<0.001$}\label{human_survey_response}
\end{figure}

Do distinct conventions also form in populations of self-play agents used to train CECP? We collected trajectories of self-play agents from the CECP partner population in example Overcooked layouts, and assessed whether the different agents would behave differently under similar scenarios. Figure \ref{cec_convention_counter_circuit} illustrates one such example in the Counter Circuit layout: When an agent is holding an onion, facing leftward, positioned right below the right pot, the population is split between policies preferring to move left, and policies preferring to move up/interact with the pot. These distinct conventions in the partner population allows the cooperator CECP agent to leverage this population diversity during training. 

To assess performance in Overcooked, we first compared our method against existing baselines using cross-play, under the 5 original layouts. For a fair comparison, prior methods (SP, FCP, MEP, PACE, CoMeDi) which do not train on a distribution of environments were separately trained on each of the 5 layouts, whereas CEC and CECP were trained on a distribution of procedurally generated layouts with the 5 original layouts held-out. 
On average, across the 5 original layouts, CECP performs significantly better than existing baselines, in spite of never having directly trained on those layouts (Figure \ref{avg_returns_combined}a, left). In an additional 100 procedurally generated heldout layouts (Figure \ref{avg_returns_combined}a, right), single-environment methods (SP, FCP, MEP, PACE, CoMeDi) fail to perform, while CECP still outperforms CEC. When pairing existing baselines and our method with human proxy agents in the original 5 Overcooked layouts, we again observed a similar result, with our method, CECP outperforming existing baselines (Figure \ref{avg_returns_combined}b). 

To directly address \textbf{Q2}, we next tested each method's ability to coordinate with humans on Coordination Ring and Counter Circuit layouts. $N=73$ participants were tasked with playing a 200 timestep game with each of the methods (FCP, MEP, CoMeDi, PACE, CEC, CECP) in randomized order, and were asked about their experiences with each method. Across both layouts, CECP again performed significantly better than all baselines in terms of average return (Figure \ref{avg_returns_combined}c). In survey responses (Figure \ref{human_survey_response}), users rated CECP above baselines in adaptability, and lower than baselines in frustration and interference.

\subsection{Next-state prediction and partner diversity shapes CECP representation}

\begin{wrapfigure}[22]{r}{0.4\textwidth}
  \centering
  \includegraphics[width=0.38\textwidth]{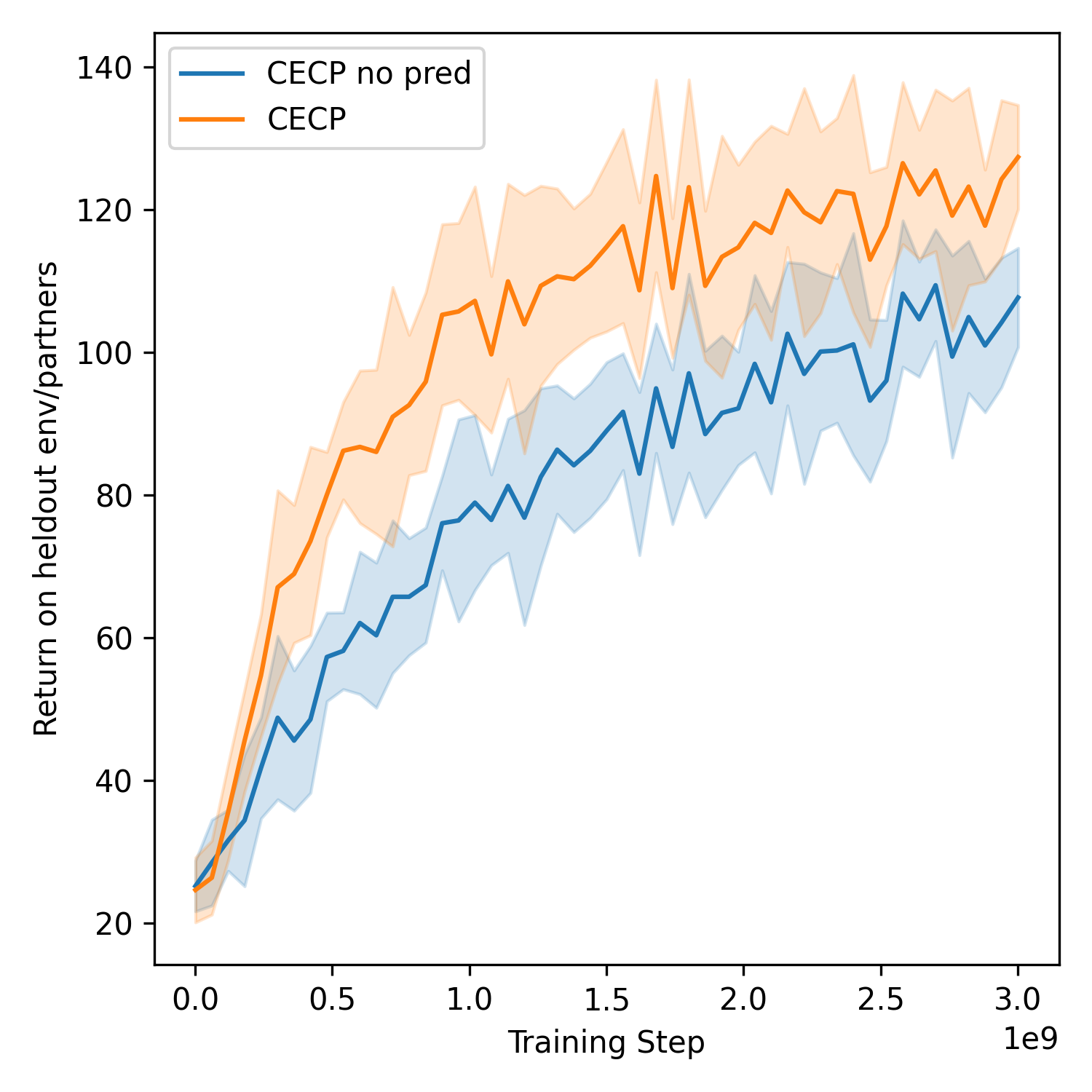}
  \caption{Comparison of CECP performance during training, with predictive next-state module (orange) vs. without (blue). Shaded regions depict standard deviation across agents.}\label{cecp_pred_ablation}
\end{wrapfigure}


How is CECP able to generalize under \SHIFT? To probe this question (\textbf{Q3}), we sought to understand which components drive CECP’s improved Human ZSC performance. Empirically, removing the next-state prediction module leads to a clear drop in cross-play performance across training (Figure \ref{cecp_pred_ablation}), pointing towards the importance of prediction. 

To better understand why, we draw on the insight of \citet{ndousse_emergent_2021}, who reason that in a cooperative task, successful next-state prediction depends not only on understanding how the environment evolves, but also on anticipating what the partner will do (see Appendix section \ref{appendix_partner_movement}, Figure \ref{state_transition_prediction}).

\begin{wrapfigure}[20]{r}{0.55\textwidth}
  \centering
    \includegraphics[width=0.55\textwidth]{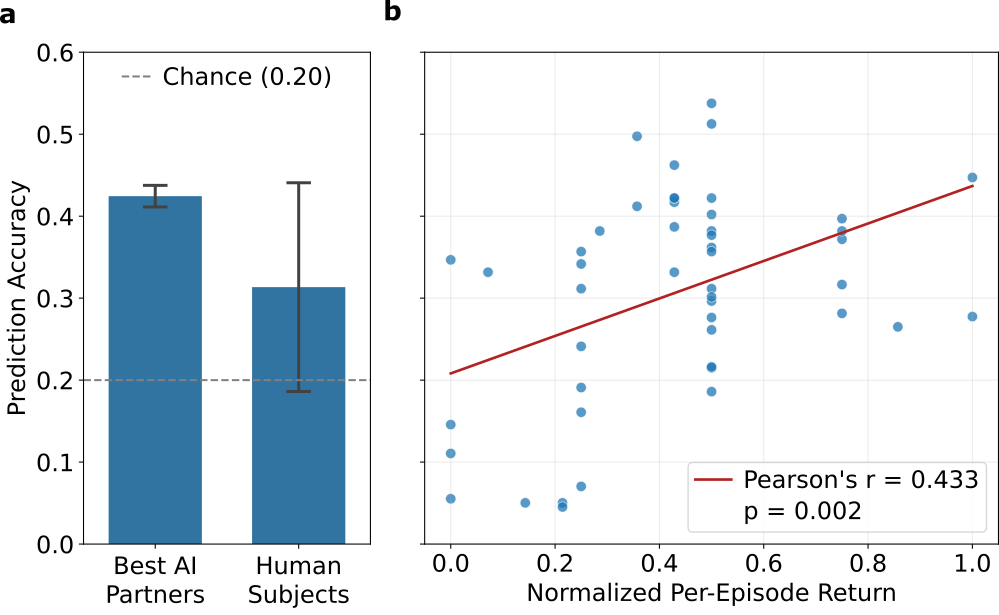}
      \caption{CECP partner movement prediction accuracy from next state predictions. \textbf{a,} prediction accuracy across episodes/timesteps for both AI training partners and human participants. \textbf{b, } correlation between CECP's prediction accuracy of human movements, and episode performance.}\label{partner_movement_acc}  
\end{wrapfigure}

More broadly, across many partners and environments, a well-trained CECP agent predicts an AI partner's movement (up, down, left, right, or stay) at rates well above chance throughout an episode (Figure\ref{partner_movement_acc}a, Best AI Partners), suggesting that next-state prediction encourages the agent to learn an implicit model of partner behavior. This movement prediction generalizes to humans as well, where human subject movements are also predicted by CECP at levels above chance (Figure \ref{partner_movement_acc}a, Human Subjects). Furthermore, we find that CECP's ability to accurately predict its human partner's movements is correlated with human's performance, suggesting that movements stemming from a more-skilled partner are easier to predict. 

\begin{figure}[t]
  \centering
  \includegraphics[width=1\textwidth]{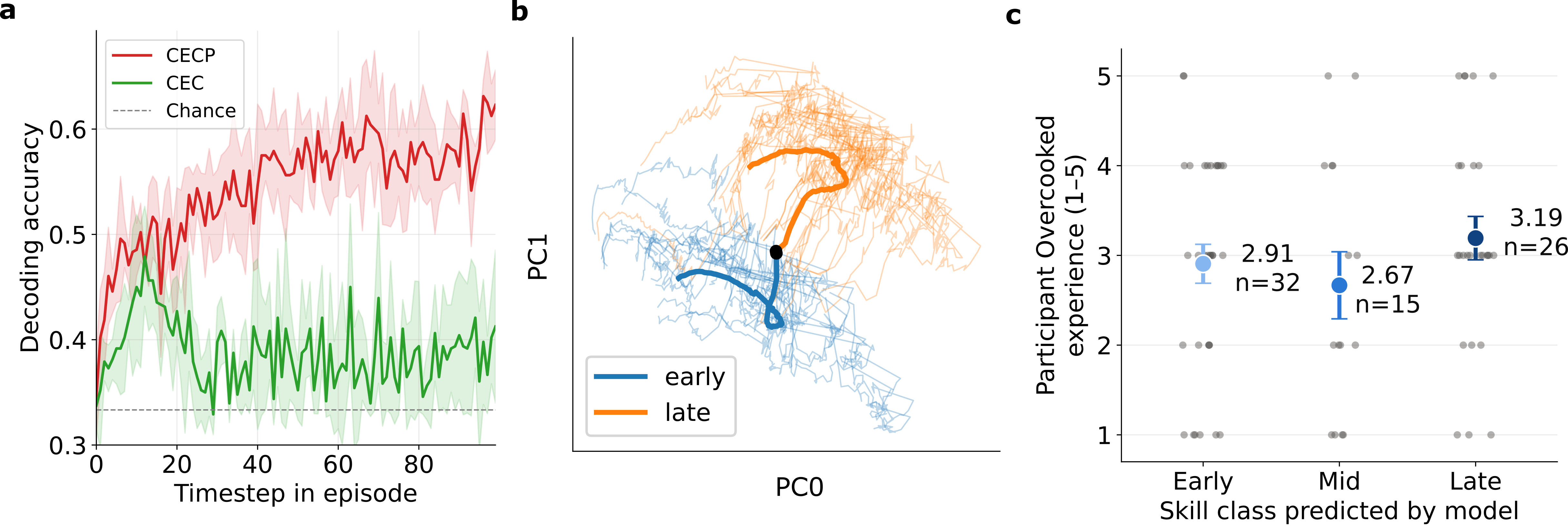}
  \caption{Skill level representation in CECP latents. \textbf{a,} prediction accuracy of partner training stage across time steps in the episode, showing the latent variable $z$ of CECP does better representation learning than the hidden layer vector of CEC. \textbf{b,} latent activity trajectories per episode visualized in principal component space, colored by partner checkpoint stage (early vs. late). Bold lines depict averaged activity trajectories per checkpoint stage, lighter lines depict 20 sub-sampled trajectories. Black dots depict activity states at the beginning of the episode. \textbf{c,} Participant's self-reported experience with Overcooked grouped by the skill level inferred by the CECP partner during interaction, using the same decoders trained in \textbf{a}. }\label{checkpoint_latent}
\end{figure}

We hypothesize that the generalization and adaptation capability of CECP comes from better inference of the partner, as the latent variable $z$ is trained from representation learning on partners with different preferences and skill levels. To investigate this, we analyze how well the latent variable $z$ can be used to predict (using a linear classifier) the partner’s training stage (early/middle/late). In an example CECP model (Figure \ref{checkpoint_latent}a, red), prediction accuracy is consistently above chance and improves as the episode unfolds. This was notably absent from prediction using the hidden layer of the CEC agent (Figure \ref{checkpoint_latent}a, green). Visualizing a CECP model's latent $z$ activity trajectories per episode in principle component (PC) space further revealed a gradual separation of latent activity based off of partner's checkpoint stage (Figure \ref{checkpoint_latent}b), further suggesting that CECP's latent space dynamically organizes around partner competency over the course of an interaction. 
When human participants interact with CECP agents, we apply the same partner training stage classifiers (early/middle/late) to the CECP agent's hidden states, and obtain a readout of the CECP agent's estimate of its human partner's skill levels. We find that the CECP agent's estimate of partner skill loosely tracks participants' self-reported Overcooked experience (Figure \ref{checkpoint_latent}c): participants classified as late-stage partners reported the highest average experience.

\begin{figure}[h]
  \centering
  \includegraphics[width=0.7\textwidth]{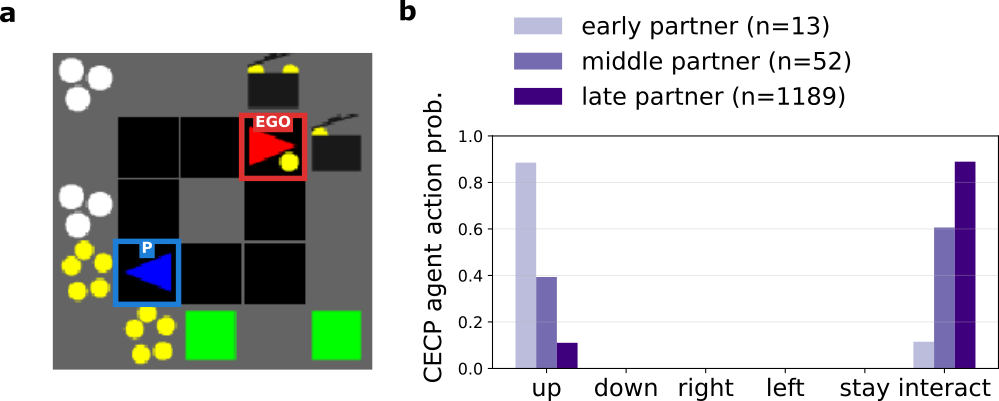}
  \caption{Averaged action distributions of an example CECP agent at a specific state in the Coordination Ring layout, with a partner in either the early, middle, or late stages of training. \textbf{a, } state illustration. CECP (ego) agent in red, partner agent in blue, at various training stages. \textbf{b, } CECP (ego) agent's averaged action distribution at this state, for each of the partner training stages. The CECP agent adjusts what action to take based on the partner's skill level, preferring to turn `up' and drop the final onion in the upper pot if the partner is less trained, in order to complete at least one soup. With more expert partner's, the CECP agent takes the more efficient action of dropping its onion in the closest available pot by using `interact'.}\label{partner_stage_coord_ring}
\end{figure}

Does inferring a partner's skill level also lead to adaptive behavior? To study this, we collected trajectories of CECP agents interacting with partner CEC agents in various training stages, identified specific states that were visited frequently with partners in each training stage, and examined the CECP agent's averaged action distributions under each of the partner training stages. Figure \ref{partner_stage_coord_ring} illustrates an example state in Coordination Ring, and highlights the CECP agent adapting its behavior based on the partner's skill level: If the partner is well trained, the CECP agent preferentially puts its onion in the right pot, as it expects the partner collect and drop an onion in the pot as well. However, if the partner is less trained, the CECP agent will prefer to move up and put its onion in the upper pot, to initiate cooking in at least one pot (Figure \ref{partner_stage_coord_ring}, see also, Appendix section \ref{do_cecp_agents_adapt}). 

\section{Conclusion} \label{conclusions_sec}

We introduce CECP, a two-stage training framework that unifies partner and environment diversity for zero-shot coordination. By training a partner population via self-play across procedurally generated environments and then training a cooperator against that population, CECP produces agents that generalize across simultaneous partner and environment shift. We further showed that a next-state prediction objective and partner diversity improves performance and encourages the cooperator to form representations that implicitly track partner behavior and skill level.

\textbf{Limitations and future work:} CECP performance is limited by the amount of diversity present in the both the partner population and environment distribution it is trained on. Future work improving on partner population diversity or broadening the range of environments may further improve generalization performance of CECP. 

We find that next-state prediction improves coordination, but its mechanism is not fully understood: the latent representation captures broad skill level rather than fine-grained, partner-specific action intent. Disentangling partner-specific and environment-specific information in the latent $z$  (for instance via a structured bottleneck) may make representations more interpretable and sample-efficient. 

Finally, while the Overcooked task has been the standard for testing Human-AI coordination, more recent benchmarks \cite{gessler_overcookedv2_2025} featuring additional complexity and partial observability may further demonstrate the efficacy of CECP. 

\begin{ack}
This research was supported by the Science of Trustworthy AI program at Schmidt Sciences, UW-Amazon Science Gift Hub, UW-Tsukuba Amazon NVIDIA Cross Pacific AI Initiative (XPAI), Tinker Research Grants, Open Philanthropy, Coefficient Giving, Toyota Research Institute Kaizen Initiative, the Schmidt AI2050 Fellows program, the NSF CISE RI program, award \#2550849, and the NSF GRFP 2023356971. This work was supported in part by Advanced Micro Devices, Inc. under the AMD University Program’s AI \& HPC Cluster. This material is based upon work supported by the Defense Advanced Research Projects Agency and the Air Force Research Laboratory, contract number(s): FA8650-23-C-7316. Any opinions, findings and conclusions, or recommendations expressed in this material are those of the author(s) and do not necessarily reflect the views of AFRL or DARPA.
\end{ack}

\newpage

\bibliographystyle{plainnat}
\bibliography{social_rl}

\newpage

\section{Appendix}

\subsection{CECP Training}

Predictor module details. (1) An RNN encoder $q_{\phi}(z \mid \tau_{0:t})$, with observations fed in via a CNN feature extractor, mapping observation histories to a latent representation $z$. (Figure \ref{intro_fig}b, green). (2) A prediction head $p_{\theta}(s_{t+1} \mid z, a_t^c)$ that decodes the next state from the latent and the cooperator's own action (Figure \ref{intro_fig}b, blue). The latent $z$ and one-hot action encoding are concatenated, projected to a flat vector, reshaped into a grid, then passed through de-convolution layers to reconstruct the next observation. (3) A policy head $\pi_{\psi}(a^c_t \mid z)$, conditioned on the latent $z$ and trained via PPO (Figure \ref{intro_fig}b, red).

Procedural generation details. To generate a distribution of environments, we leverage the procedural environment generation method proposed in \cite{jha_cross-environment_2025}, varying onion, pot, and delivery port placement as well as layout orientation, which is an extension of the JaxMARL project \cite{rutherford_jaxmarl_2024}.

CECP training first involves training a population of self-play agents across a distribution of environments. To do so, we follow the procedural environment generation and training proposed in CEC \cite{jha_cross-environment_2025}. 

To train CECP cooperator agents, 6 trained partner agents. along with their early, middle, late training checkpoints are used to form a partner population, with 2 partner agents held-out for testing. Cooperator and partner agents were trained on a cluster of NVIDIA A100 GPUs. Each step took around 1 day of training. 
The following hyperparameters were used to train the CECP cooperator agent: 

\begin{table}[H]
\centering
\begin{tabular}{cc}
\hline
hyperparameter & value \\
\hline
CNN kernels & [2, 2], [2, 2] \\
CNN channels & [64, 32] \\
encoder FC sizes & [512, 512] \\
readout size & 256 \\
recurrent layer type & LSTM \\
recurrent layer size & 256 \\
recurrent layers & 1 \\
actor head sizes & [256, 192, 128, 64] \\
critic head sizes & [512, 256, 192, 128] \\
decoder projection size & $9 \times 9 \times 32$ \\
deconv channels & [64, 26] \\
deconv kernels & [2, 2], [2, 2] \\
deconv strides & [1, 1] \\
environment steps & 3B (per agent) \\
parallel environments & 4096 (per agent) \\
episode length & 256 \\
PPO batch size & 16384 \\
PPO epochs & 4 \\
PPO learning rate & 0.0003 \\
LR schedule & linear \\
Generalized Advantage Estimator (GAE) $\lambda$ & 0.95 \\
discounting factor $\gamma$ & 0.99 \\
entropy coefficient & 0.005 \\
ratio clip & 0.2 \\
gradient clip norm & 0.5 \\
predictive loss coefficient & 1000 \\
\hline
\end{tabular}
\caption{Hyperparameters for CECP Cooperator}
\end{table}

\subsection{Cross-play comparison between CEC and CECP agents in Cooperative Foraging}

\begin{figure}[H]
  \centering
  \includegraphics[width=0.8\textwidth]{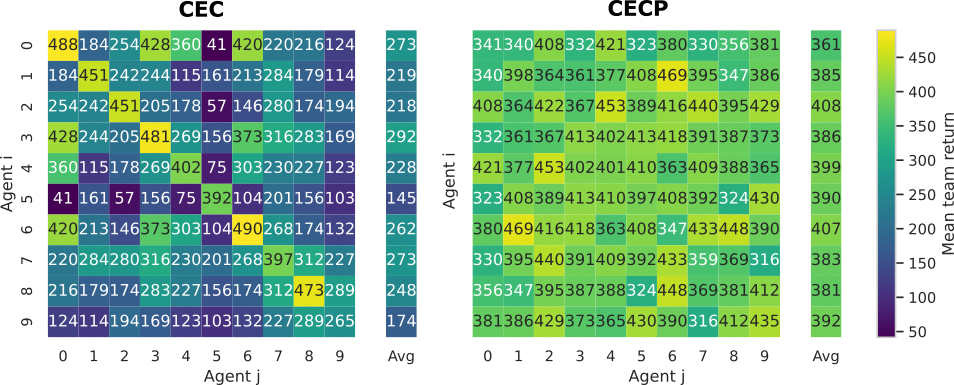}
  \caption{Comparison of agent-by-agent cross-play returns of CEC agents (left) and CECP agents (right) on the cooperative foraging task, on held-out goal locations. Agents in each heatmap are trained with the same method, but different initial seeds. Diagonals depict self-play results. The reduced gap between diagonal and off-diagonal items shows CECP improves robustness to novel partners compared to CEC.}
  \label{cp_cross_play_compare}
\end{figure}

\subsection{Assessing behavioral diversity of partner populations}

Our method CECP, along with many of the baselines we include (FCP, MEP, CoMeDi), is a population-based training method, so we sought to quantify the relative behavioral diversity within each methods' partner populations. 

In each of the original 5 Overcooked layouts, for a population of $N=8$ agents $\{\pi_1, \dots, \pi_N\}$, we first collect a set of trajectories $\mathcal{D}$ by pooling each agent's self-play trajectories across a fixed number of episodes. For each agent $\pi_j$, we replay trajectories from $\mathcal{D}$, and compute the agent's average action distribution at each unique state $s$, over the $|\mathcal{A}| = 6$ actions: $\pi_j(\cdot \mid s)$. We then compute the population Jensen-Shannon Divergence (JSD), averaged across states: 

\begin{equation}
\mathrm{JSD}_{\mathrm{pop}} = \mathbb{E}_{s \sim \mathcal{D}} \Big[ H\big(\bar{\pi}(\cdot \mid s)\big) - \frac{1}{N} \sum_{j=1}^{N} H\big(\pi_j(\cdot \mid s)\big) \Big],
\end{equation}

where $\bar{\pi}(\cdot \mid s) = \frac{1}{N}\sum_{j=1}^{N} \pi_j(\cdot \mid s)$ is the average of the population's action distributions at $s$, and $H$ the Shannon entropy. 

While $\mathrm{JSD}_{\mathrm{pop}}$ gives an overall assessment of population diversity, we are further interested in how similar a pair of agents in the population could be, in the worst case scenario. For all pairs of agents $(i, j)$ in a population, we find the minimum state-averaged pairwise Jenson-Shannon Divergence: $\min \mathrm{JSD}_\text{pair} = \min_{i \neq j} \mathrm{JSD}_{ij}$. In Table \ref{tab:oc_population_diversity}, we report $\mathrm{JSD}_\text{pop}$, and $\min \mathrm{JSD}_\text{pair}$ in parentheses, for the partner populations of each method, assessed in each of the original 5 Overcooked layouts. 

\begin{table}[H]
\centering
\caption{Diversity of each method's partner populations on Overcooked. Each cell reads $\mathrm{JSD}_\text{pop}$ ($\min \mathrm{JSD}_\text{pair}$). Populations of 8 agents per method.}
\label{tab:oc_population_diversity}
\small
\begin{tabular}{lcccc}
\toprule
Layout & \textbf{CECP} & \textbf{FCP} & \textbf{MEP} & \textbf{CoMeDi} \\
\midrule
Asymmetric Advantages & 0.387 (0.151) & 0.242 (0.067) & 0.370 (0.069) & 0.389 (0.079) \\
Coordination Ring & 0.245 (0.089) & 0.384 (0.152) & 0.409 (0.124) & 0.456 (0.169) \\
Counter Circuit & 0.334 (0.131) & 0.377 (0.137) & 0.366 (0.136) & 0.493 (0.179) \\
Cramped Room & 0.222 (0.093) & 0.261 (0.104) & 0.336 (0.120) & 0.467 (0.155) \\
Forced Coordination & 0.308 (0.127) & 0.281 (0.112) & 0.200 (0.005) & 0.342 (0.088) \\
\midrule
All layouts & 0.299 (0.089) & 0.309 (0.067) & 0.336 (0.005) & 0.429 (0.079) \\
\bottomrule
\end{tabular}
\end{table}

Overall, we observe that the CECP partner population’s behavioral diversity is similar to that of FCP's partner population, though across all layouts, CECP’s $\min \mathrm{JSD}_\text{pair}$ is slightly higher than other methods. While diversity-aware methods (MEP, CoMeDi) increase $\mathrm{JSD}_\text{pop}$, they also produce lower $\min \mathrm{JSD}_\text{pair}$ in certain layouts, suggesting that their policies end up converging to a few specific conventions. 

\begin{figure}[H]
  \centering
  \includegraphics[width=0.8\textwidth]{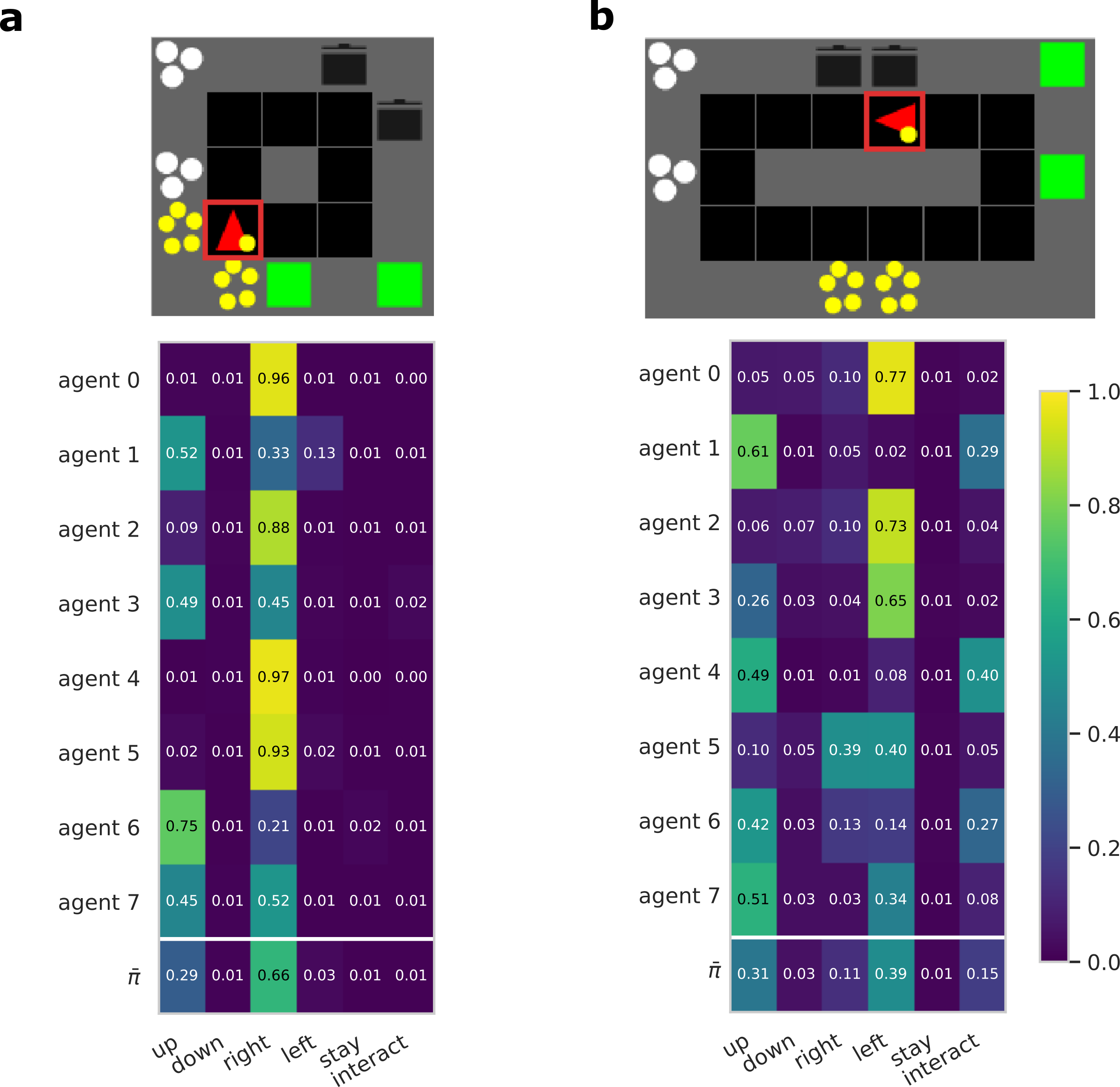}
  \caption{Two example scenarios where CECP's partner population exhibits distinct conventions, for the Coordination Ring (a) and Counter Circuit (b) layouts. Top, the position, orientation, and item status of the agents in interest. Bottom, the the average action distributions of each agent in the partner population.}\label{cec_convention}
\end{figure}

\begin{figure}[H]
  \centering
  \includegraphics[width=1\textwidth]{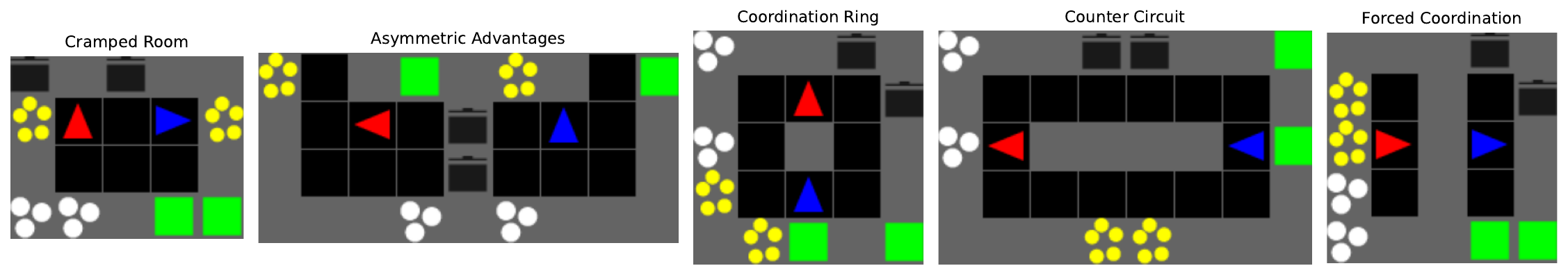}
  \caption{Visualization of the original 5 Overcooked environments. In Overcooked, two chefs need to cooperate together to make and deliver the soups efficiently.} \label{oc_original_5_layouts}
\end{figure}

To further confirm that agents in CECP's partner population are indeed forming distinct conventions, we next examined each partner's average action distributions under specific scenarios in the Coordination Ring and Counter Circuit layouts. In Coordination Ring, when an agent is holding an onion, facing upward, positioned in the bottom left corner, the population is split between policies preferring to move up and initiate a clockwise movement, and policies preferring to moving right to initiate a counterclockwise movement (Figure \ref{cec_convention}a). Similarly, in Counter Circuit, when an agent is holding an onion, facing leftward, positioned right below the right pot, the population is split between policies preferring to moving left, and policies preferring to move up/interact with the pot (Figure \ref{cec_convention}b).

\subsection{Training human proxy agents}

\paragraph{Human dataset} We use the human dataset released by~\citet{carroll_utility_2020} for five Overcooked layouts (Asymmetric Advantages, Coordination Ring, Counter Circuit, Cramped Room, and Forced Coordination) and the same data processing and behavior cloning (BC) implementations as~\citet{liang_learning_2024}.

The human dataset is split into a training split and a test split with a ratio of 70\% / 30\%. The average episode length is $T\approx 1200$. The statistics of the training/test splits are shown in the following table.

\begin{table}[H]
\centering
\small
\begin{tabular}{lcc}
\toprule
 & \multicolumn{2}{c}{Trials}  \\
\cmidrule(lr){2-3}
Layout & Train & Test \\
\midrule
Asymmetric Advantages & 14 & 5 \\
Coordination Ring & 13 & 5 \\
Counter Circuit & 13 & 5 \\
Cramped Room & 14 & 5 \\
Forced Coordination & 9 & 3 \\
\bottomrule
\end{tabular}
\end{table}

For each layout, we train two BC models on the training and test splits, respectively. The held-out BC model trained on the test split is used for evaluation only and is never used in training. The hyperparameters for training are shown in the following table.

\begin{table}[H]
\centering
\begin{tabular}{cc}
\hline
hyperparameter & value \\
\hline
hidden layer sizes & [64, 64] \\
number of MLP layers & 2 \\
learning rate & 0.001 \\
batch size & 64 \\
training epochs & 100 \\
\hline
\end{tabular}
\caption{Hyperparameters for BC human proxy models}
\end{table}

\subsection{Human-AI Coordination Evaluation}

\textbf{Participants:} We recruited $N = 73$ participants via Prolific under a protocol approved by our university's IRB. Participants were aged 19–72 (median = 28.5), of whom 43.1\% identified as male, 55.6\% as female, and 1.4\% did not to say.
Prior to the experiment, participants reported their familiarity with the Overcooked video game on a 5-point Likert scale (1 = no experience, 5 = very experienced). The median experiment time was 34 minutes, and each participant was paid \$11 USD (averaging \$19.46/hr) for their time and effort. 

\textbf{Procedure:} Following a tutorial round, each participant was randomly assigned to one of two layouts: Counter Circuit or Coordination Ring. Participants then played one episode with each of the tested AI methods, for a total of 6 episodes per participant. For each method, a random agent was drawn from a pool of 8 trained agents (each initialized with a different random seed) to control for seed-level variance. Each episode lasted 200 timesteps.

\textbf{Post-episode survey:} After each episode, participants rated their experience on the following statements \cite{jha_cross-environment_2025} using a 5-point Likert scale (1 = strongly disagree, 5 = strongly agree):
\begin{itemize}
    \item The agent adapted to me when making decisions.
    \item The agent was consistent in its actions.
    \item The agent's actions were human-like.
    \item The agent frequently got in my way.
    \item The agent's behavior was frustrating.
    \item Overall, I enjoyed playing with the agent.
    \item Overall, I felt that the agent's ability to coordinate with me was...
\end{itemize}

\subsection{Overcooked evaluation results by layout}

\begin{figure}[H]
  \centering
  \includegraphics[width=1\textwidth]{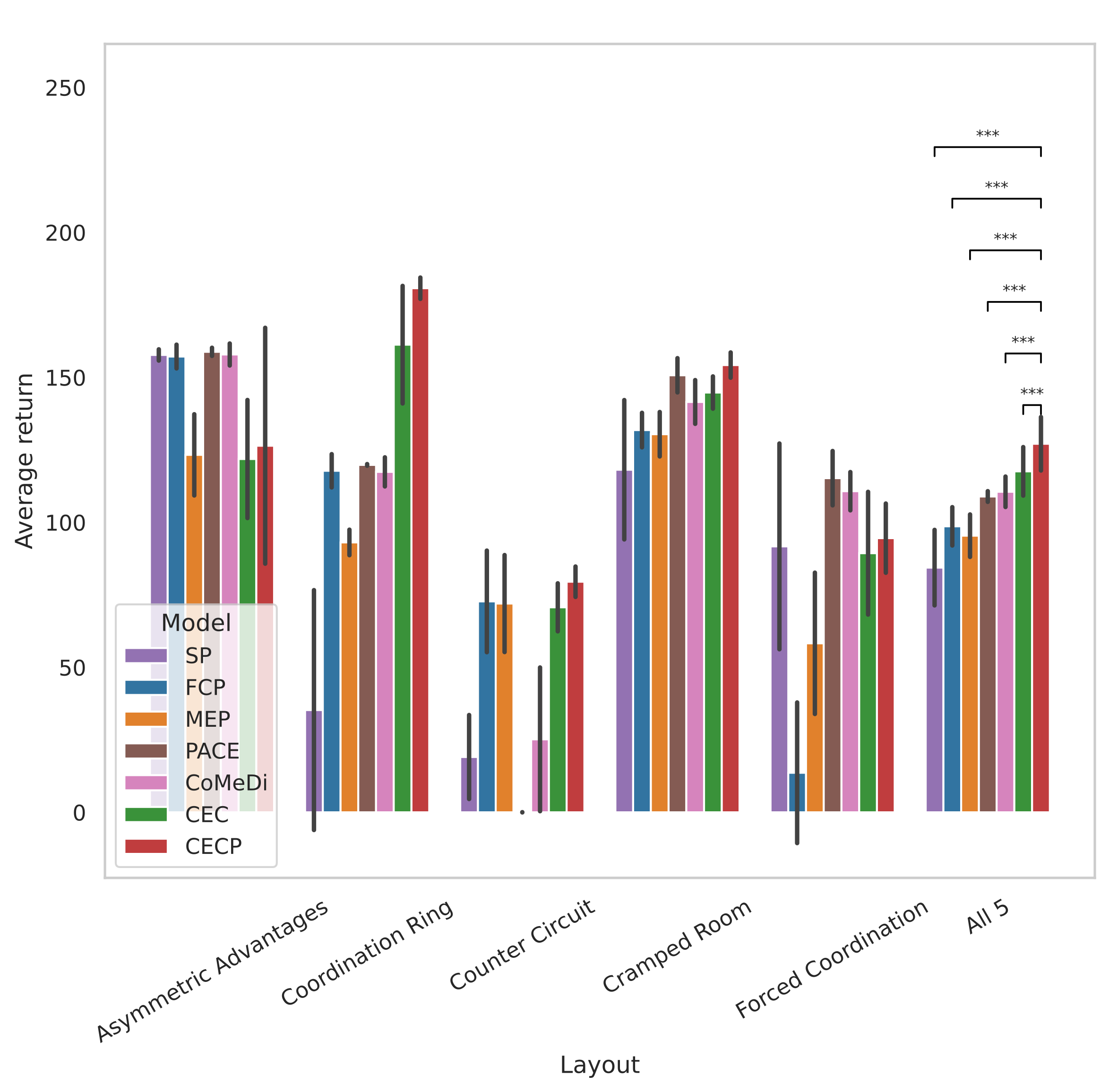}
  \caption{Cross-play average returns split by the original 5 Overcooked layouts} \label{oc_cross_play_original_5}
\end{figure}

\begin{figure}[H]
  \centering
  \includegraphics[width=1\textwidth]{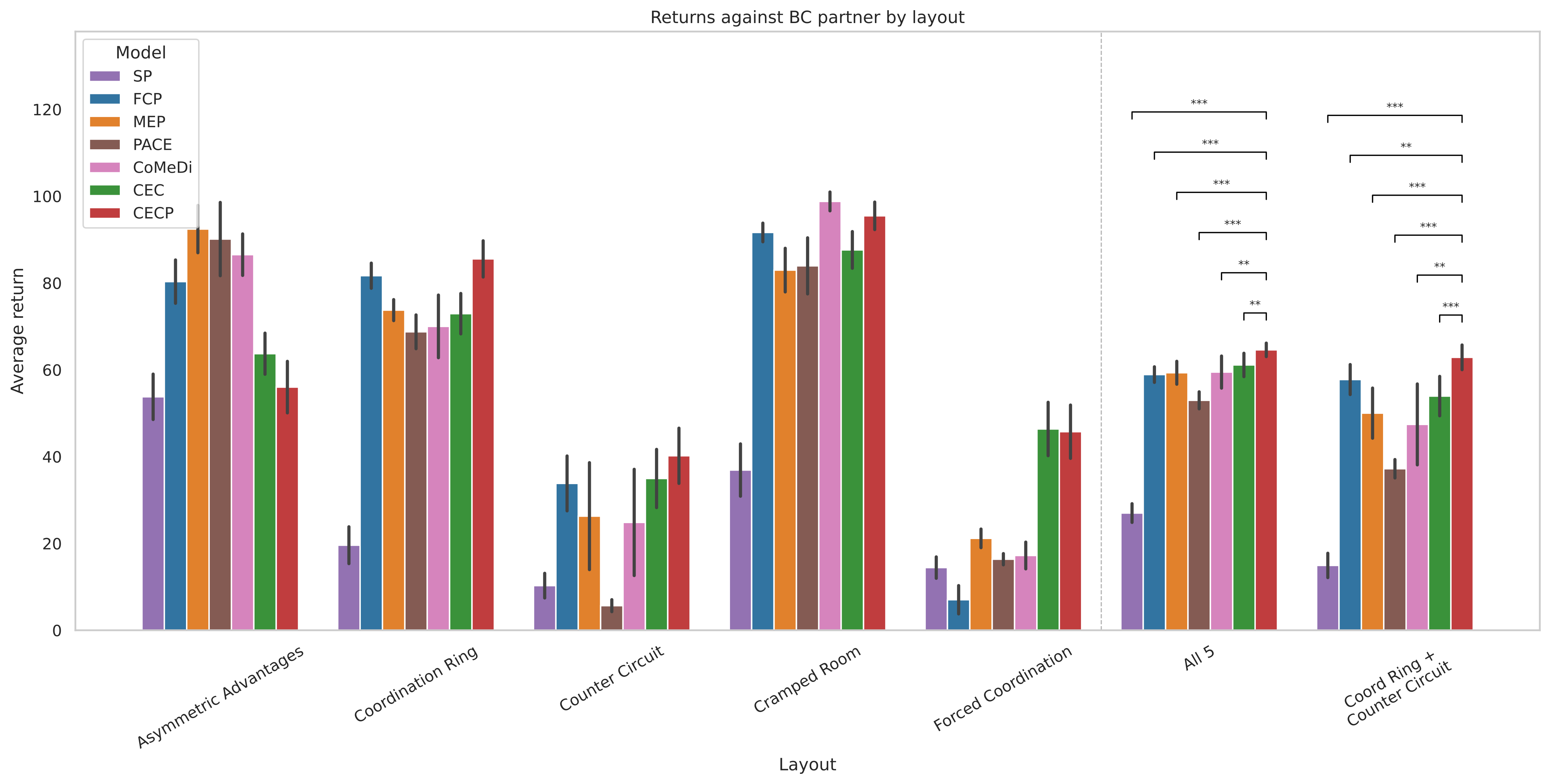}
  \caption{Average returns when partnering with human proxies, split by the original 5 Overcooked layouts} \label{oc_bc_original_5}
\end{figure}

\subsection{Partner movement prediction}\label{appendix_partner_movement}
To assess whether agents implicitly model their partner's behavior, we evaluate partner action prediction accuracy as a proxy for learned partner representations. At each timestep, the CECP cooperator agent is queried for its predicted next state conditioned on its own sampled action. The partner's predicted action is then inferred from this predicted state by examining the difference in partner positions between the predicted next state and the current state. 

\begin{figure}[H]
  \centering
  \includegraphics[width=0.8\textwidth]{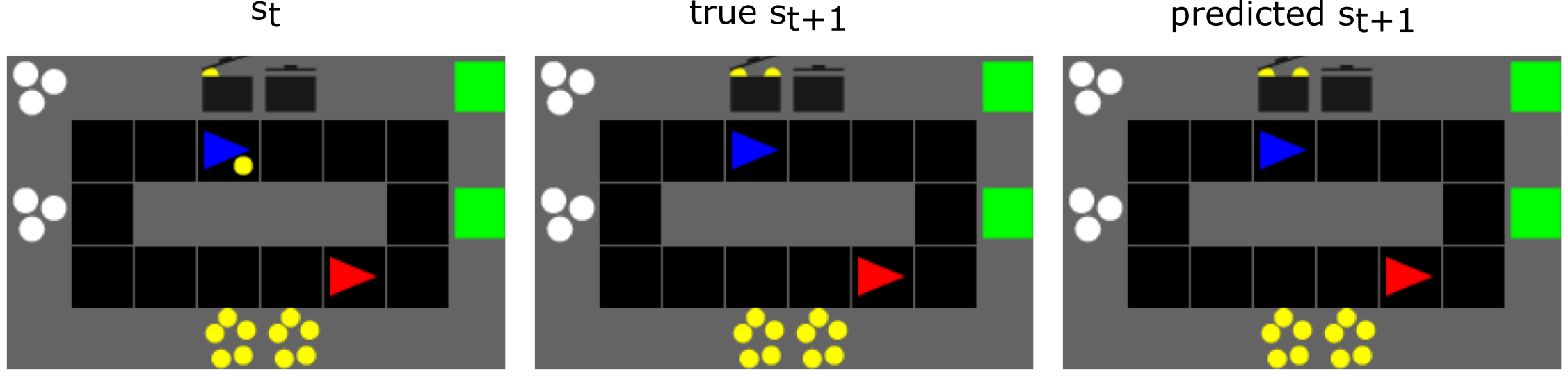}
  \caption{Example next-state prediction from a CECP agent (red), coordinating with a partner (blue). At state $s_t$ (left), in order to fully predict the true next state $s_{t+1}$ (middle), CECP's prediction (right) must capture both the blue partner's action to put the onion in the pot, and the environment transition of onions in the pot increasing by 1. }\label{state_transition_prediction}
\end{figure}

\subsection{Latent Representation Analysis}

\textbf{Partner Checkpoint decoding:} We evaluate a trained CECP agent that is paired with partners of early/middle/late checkpoint stages. For each partner, we run 20 episodes, and collect the per-timestep activations in the CECP agent’s latent $z$ layer . At each timestep, a multi-class linear discriminant analysis LDA classifier is fit across partners and episodes on the full latent vector to classify checkpoint stage. We report the average test classifier accuracy using 5-fold cross-validation. 

\textbf{PCA analysis:} The matrix collected from a CECP agent’s latent z activity is reshaped to $(\text{n\_units}, \text{n\_timesteps} \times \text{n\_episodes})$, mean-centered per unit, and decomposed via full SVD. Episode trajectories are projected into the space of the top two principle components, and plotted as time-indexed curves colored by partner label (early vs. late), with average trajectories plotted as well. 

\subsection{Do CECP agents adapt their policies to partners at different training stages?}\label{do_cecp_agents_adapt}

In Figure \ref{checkpoint_latent}, we observed that partner training stage is decodable from the latent layer of a CECP agent. Here, we additionally assess whether a CECP agent adapts its policy to partners of different training stages. Let $A$ denote the CECP agent's action, $Z \in \{\text{early}, \text{middle}, \text{late}\}$ the training stage of the partner's checkpoint, and $S$ the environment state. We collected 8000 episode trajectories of CECP agents partnering with partners from each training stage $z$, and selected states $s$ that the CECP agent had visited more than 10 times with partners of every stage. To further ensure that the CECP agent had sufficient time to infer the partner training stage, only states that were visited after the first 10 steps of an episode were considered. For each such state and partner stage, we computed the CECP agent's averaged action distribution $\pi(a \mid s, z)$, and from these the conditional mutual information between the agent's action and the partner stage, treating $Z$ as uniform over the three stages:

\begin{equation}
I(A; Z \mid S = s) = \frac{1}{3} \sum_{z} \mathrm{KL}\left( \pi(a \mid s, z) ,\middle|, \bar{\pi}(a \mid s) \right), \qquad \bar{\pi}(a \mid s) = \frac{1}{3} \sum_{z} \pi(a \mid s, z).
\end{equation}

\begin{figure}[H]
  \centering
  \includegraphics[width=0.8\textwidth]{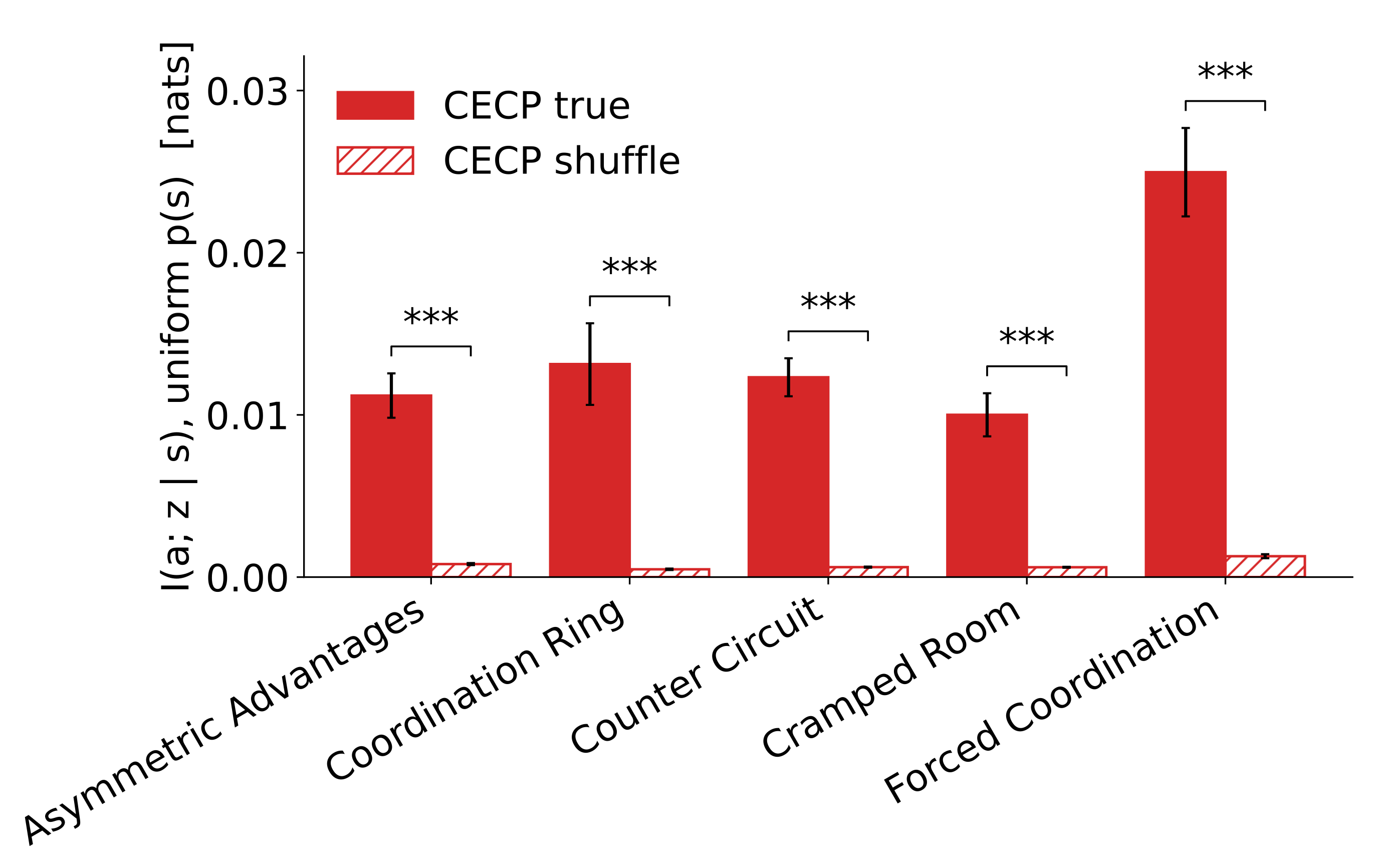}
  \caption{Average mutual information between a CECP agent's action and the partner's training stage, averaged uniformly over selected states, for each of the five original Overcooked layouts. Solid bars: true partner-stage labels. Hatched bars: shuffled null. Bars show the mean over the 8 CECP agents of the population; error bars are SEM. $^{***}$: $p < 0.001$ by permutation test; for every agent on every layout}\label{partner_stage_mi}
\end{figure}

For each of the original 5 Overcooked layouts, the mutual information averaged uniformly across selected states is significantly above a shuffled null distribution, generated by randomly permuting partner training stage labels across trajectories and recomputing the averaged mutual information, repeated 1000 times (Figure \ref{partner_stage_mi}). 

\begin{figure}[H]
  \centering
  \includegraphics[width=0.8\textwidth]{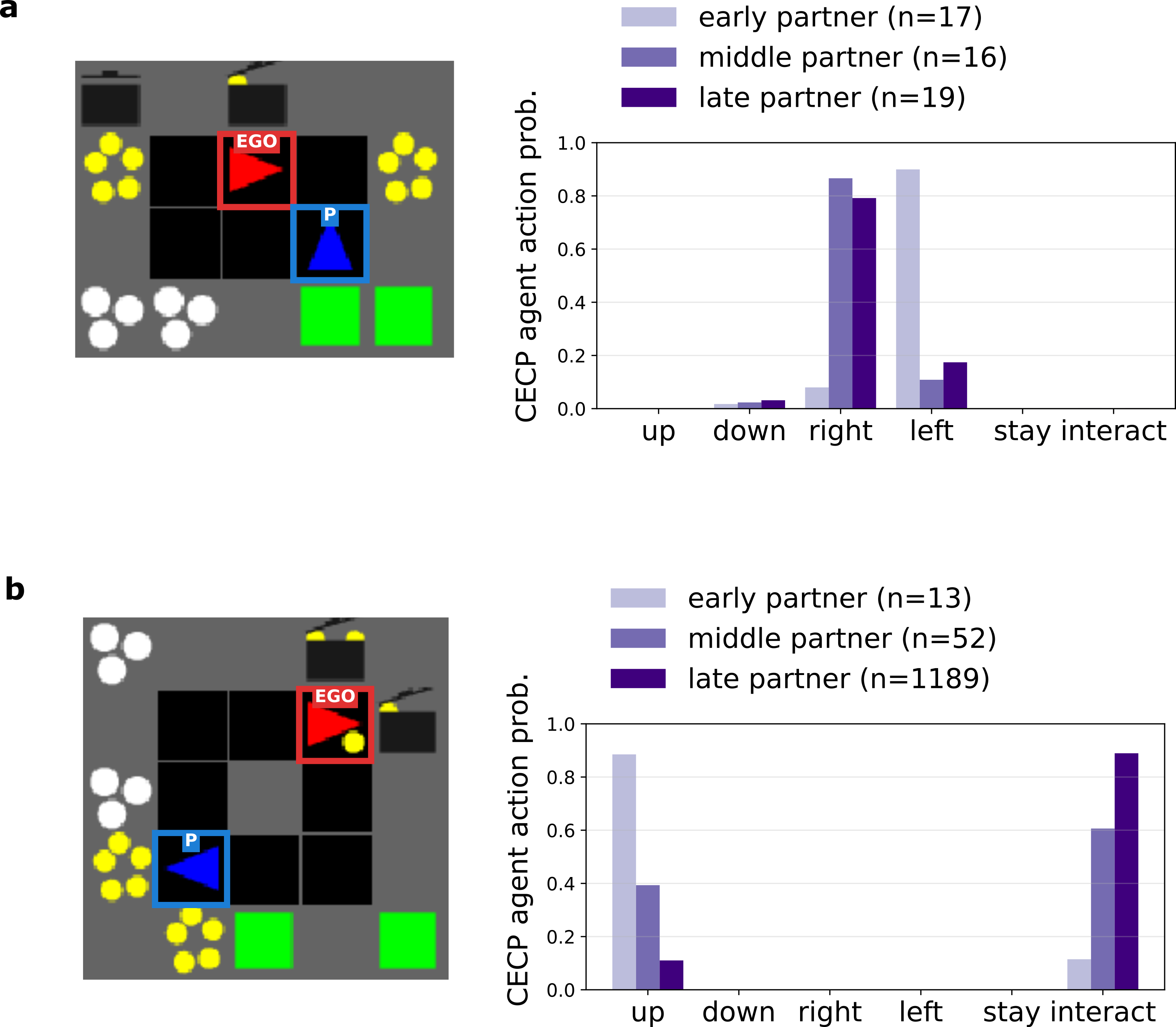}
  \caption{Averaged action distributions of example CECP agents at specific states, when the partner is in the early/middle/late stages of training. Example states for Cramped Room (a) and Coordination Ring (b). Left, state illustration, CECP (ego) agent in red, partner agent in blue, at various training stages. Right, CECP (ego) agent's averaged action distribution at this state, for each of the partner training stages. }\label{partner_stage_examples}
\end{figure}

When we examined specific states where mutual information was high, it became more clear that CECP agents were adapting their behavioral strategies according to the partner's training stage. For example, Figure \ref{partner_stage_examples}a illustrates a CECP agent (in red) in Cramped Room that will tend to avoid the partner (in blue) (move more left since the partner is to the right) if the partner is the early stages of training. Additionally, Figure \ref{partner_stage_examples}b illustrates a specific scenario in Coordination Ring, where the top pot already has two onions, and the right pot has one. If the partner is well trained, the CECP agent preferentially puts its onion in the right pot, as it expects the partner collect and drop an onion in the pot as well. However, if the partner is less trained, the CECP agent will prefer to move up and put its onion in the upper pot, to initiate cooking. 

\subsection{LLM Usage}
We leveraged Claude Code to aide in the implementation of our methods and analyses, and to help with the writing/editing of this manuscript. 


\end{document}